\documentclass[conference,a4paper]{IEEEtran}
\IEEEoverridecommandlockouts
\usepackage{cite}
\usepackage{amsmath,amssymb,amsfonts}
\usepackage{textcomp}
\usepackage[table,xcdraw]{xcolor}
\usepackage{multirow}
\usepackage{acronym} 
\usepackage{svg}
\usepackage[hypcap=true]{subcaption, caption}
\usepackage{graphicx}

\usepackage[citecolor=black,
            colorlinks=true,
            linkcolor=black,
            urlcolor  = black,
            pdfpagelabels={true},
            pdftitle={Receptive Field Analysis of Temporal Convolutional Networks for Monaural Speech Dereverberation},
            pdfauthor={William Ravenscroft, Stefan Goetze, Thomas Hain},
            pdfkeywords={speech, dereverberation, temporal convolutional network, enhancement, sequence modelling, tasnet},
            pdfsubject={speech, dereverberation, temporal convolutional network, enhancement, sequence modelling, tasnet}]
            {hyperref}
\usepackage{orcidlink}
\acrodef{AC}    {acoustic conditions}
\acrodef{ASR}   {automatic speech recognition}
\acrodef{BLSTM} {bidirectional long short term memory}
\acrodef{CB}    {convolutional block}
\acrodefplural{CB}[CBs]{convolutional blocks}
\acrodef{Conv-TasNet} {convolutional \ac{TasNet}}
\acrodef{CSM}   {clean speech mixtures}
\acrodef{cLN}   {channelwise layer normalization}
\acrodef{DAE}   {denoising autoencoder}
\acrodef{DL}    {deep learning}
\acrodef{DNN}   {deep neural network}
\acrodef{DPRNN} {dual path recurrent neural network model}
\acrodef{DPTNet}{dual path Transformer network}
\acrodef{E2E}   {end-to-end}
\acrodef{ER}    {early reflection}
\acrodefplural{ER}[ERs]{early reflections}
\acrodef{gLN}   {global layer normalization}
\acrodef{LSTM}  {long short term memory}
\acrodef{LR}    {late reflection}
\acrodefplural{LR}[LRs]{late reflections}
\acrodef{MHA}   {Multihead Attention}
\acrodef{MOS}   {Mean Opinion Score}
\acrodef{MR}    {mask refinement}
\acrodef{NMF}   {non-negative matrix factorization}
\acrodef{NSM}   {noisy speech mixture}
\acrodef{NRSM}  {noisy reverberant speech mixture}
\acrodef{PESQ}  {perceptual evaluation of speech quality}
\acrodef{PIT}   {permutation invariant training}
\acrodef{PM}    {post-masking}
\acrodef{PReLU} {parametric \ac{ReLU}}
\acrodef{ReLU}  {rectified linear unit}
\acrodef{RF}    {receptive field}
\acrodef{RIR}   {room impulse response}
\acrodef{RIRs}   {room impulse responses}
\acrodefplural{RIR}[RIRs]{room impulse responses}
\acrodef{RSM}   {reverberant speech mixture}
\acrodef{SISDR} {scale-invariant signal-to-distortion ratio}
\acrodef{SDR}   {signal-to-distortion ratio}
\acrodef{SNR}   {signal-to-noise ratio}
\acrodef{SRMR}  {speech-to-reverberation modulation energy ratio}
\acrodef{SP}    {signal processing}
\acrodef{STFT}  {short-time Fourier transform}
\acrodef{STOI}  {short-time objective intelligibility}
\acrodef{ESTOI} {extended short-time objective intelligibility}
\acrodef{TasNet}{Time-domain audio separation network}
\acrodef{TCN}   {temporal convolutional network}
\acrodef{UPGMA} {unweighted pair group method with arithmetic mean}
\acrodef{WER}   {word error rate}
\acrodef{WPE}   {weighted prediction error}

\newcommand{\mat}[1]{\mathbf{#1}}
\newcommand{\vekt}[1]{\ensuremath{\boldsymbol{\mathrm{#1}}}}

\newcommand{\Real}{\mathbb{R}}

\def\BibTeX{{\rm B\kern-.05em{\sc i\kern-.025em b}\kern-.08em
    T\kern-.1667em\lower.7ex\hbox{E}\kern-.125emX}}

\begin{document}
\title{Receptive Field Analysis of Temporal Convolutional Networks for Monaural Speech Dereverberation\\
\thanks{{This work was supported by the Centre for Doctoral Training in Speech and Language Technologies (SLT) and their Applications funded by UK Research and Innovation [grant number EP/S023062/1].} {This work was also funded in part by 3M Health Information Systems, Inc.}}
}

\author{\IEEEauthorblockN{William Ravenscroft$^{\orcidlink{0000-0002-0780-3303}}$, Stefan Goetze$^{\orcidlink{0000-0003-1044-7343}}$, and Thomas Hain$^{\orcidlink{0000-0003-0939-3464}}$}

\IEEEauthorblockA{\textit{Department of Computer Science}, 
\textit{{The} University of Sheffield}, Sheffield, United Kingdom \\
\{jwravenscroft1, s.goetze, t.hain\}@sheffield.ac.uk\vspace*{-0.2cm}}
}

\maketitle

\begin{abstract}
Speech dereverberation is often an important requirement in robust speech processing tasks. Supervised \ac{DL} models give 
{state-of-the-art} performance for 
{single-channel } speech dereverberation. 
{\Acp{TCN} are}
commonly used for sequence modelling in speech enhancement tasks. A feature of \ac{TCN}s is that they have a \ac{RF} dependent on the specific model configuration which determines the number of input frames that can be observed to produce an individual output frame. It has been shown that \ac{TCN}s are capable of performing dereverberation of simulated speech data, however a thorough analysis{, especially with focus on the \ac{RF} is yet lacking in the literature.}
This paper 
{analyses dereverberation performance depending on} 
the model size and the \ac{RF} of 
{\acp{TCN}.}
{Experiments} using the WHAMR corpus which is extended to include \acp{RIR} with larger T60 values
{demonstrate} 
that a larger \ac{RF} can have significant improvement in performance when training smaller \ac{TCN} models.
It is also demonstrated that \ac{TCN}s benefit from a wider \ac{RF} when dereverberating \acp{RIR} with larger RT60 values. 
\end{abstract}


\section{Introduction}
In 
{far-field} recording environments, reverberation affects the quality and intelligibility of the recorded audio signal {\cite{NG10}}. This remains a problem for many domains in speech technology
\cite{FFASRHaebUmbach,Hain}. Dereverberation of speech signals has been studied thoroughly over the past {decades}
\cite{Cauchi_REVERB_2015,1161990}
{based on machine}
learning models and \ac{SP} techniques 
\cite{FFASRHaebUmbach,habets2007single}.

Most \ac{SP} approaches model reverberant speech 
as a mixture of the {anechoic} speech signal summed with delayed{,} exponentially decaying weighted sums of itself. The sequence of weights used in this summation is commonly referred to as the \ac{RIR}{, which }
is typically modelled in three parts: the direct path, the \acp{ER} and the \acp{LR} \cite{habets2007single}. 
\Acp{ER} in speech are typically assumed to occur within the first {50~ms after}
the direct path.
\ac{SP} methodologies for 
suppressing reverberant content in speech signals range from a number of techniques with the most prominent approaches in recent work using {spectral suppression or} linear predictive modelling \cite{Park2018,Zhang2020OnEM}. 

\ac{DL} models have mostly surpassed pure \ac{SP} approaches for enhancing reverberant speech signals on objective measures such as 
\ac{WER} 
{or} \ac{PESQ} \cite{Purushothaman,Kinoshita2017,9095210}. 
{\Acp{TasNet} \cite{tasnet} were} proposed for speech separation which 
{were} later applied also to dereverberation 
\cite{reverbtasnet}. 
{\Acp{Conv-TasNet} \cite{convtasnet} replace} 
the BLSTM network of 
{\acp{TasNet}}
with a fully convolutional model using a \ac{TCN} \cite{tcn}. \acp{TCN} have also shown to be effective at more general speech enhancement tasks including dereverberation \cite{WHAMR}. A dereverberation network using a \ac{TCN} with self attention was proposed {in} \cite{9095210} which demonstrated {that} \ac{TCN} models give competitive results with other state-of-the-art techniques such as \ac{DNN} \ac{WPE} models.

In this work, 
{\acp{Conv-TasNet} are analysed for} 
application to monaural dereverbation of speech. 
{The main}
focus 
is to 
{analyse}
the interplay between \ac{RF}, model size and \ac{RIR} length on the capability of Conv-TasNets to dereverb speech.

The remainder of the paper proceeds as follows, Section~\ref{sec:2} describes the signal model, in Section~\ref{sec:3} \ac{Conv-TasNet} is formulated as a \ac{DAE}, in Section~\ref{sec:4} the data and experimental setup are discussed, in Section~\ref{sec:5} 
results are presented and 
Section~\ref{sec:6} {concludes the paper.}

\section{Signal Model}\label{sec:2}
A discrete single channel reverberant speech signal
\begin{equation} \label{eq:2:2}
    x[i] = h[i]\ast s[i] = s_\mathrm{dir}[i] + {s_\mathrm{rev}[i]}
\end{equation}
for discrete time index $i$ can be decomposed into direct-path signal $s_\mathrm{dir}[i]=$ $\alpha s[i-i_0]$, with a delay $i_0$ and possible attenuation by a factor $\alpha$, and reverberant part $s_\mathrm{rev}[i]$. In \eqref{eq:2:2}, $h[i]$ is the \ac{RIR} and $\ast$ denotes the convolution operation. The signal of length $L_s$ can be split into  $L_{\vekt{x}}$ blocks of length $L_{\mathrm{BL}}$ with a 50\% overlap and block index $\ell$ defined as
\begin{equation}\label{eq:InputSignalBlock}
    \vekt{x}_\ell = 
\left[x[0.5(\ell-1)L_{\mathrm{BL}}],\ldots, x[0.5(1+\ell) L_{\mathrm{BL}}-1]\right]
\end{equation}
The aim of this paper is {to} estimate the values of ${\vekt{s}}_\mathrm{dir}=[{s}_\mathrm{dir}[0],\ldots,{s}_\mathrm{dir}[L_s-1]]$ denoted as ${\hat{\vekt{s}}=[\hat{s}[0],\ldots,\hat{s}[L_s-1]]}$.

\section{Dereverberation Network}\label{sec:3}
The dereverberation network is based on reformulating \ac{Conv-TasNet} 
as a \ac{DAE} composed of an encoder, {a} mask estimation network and a decoder \cite{convtasnet,RGH22_AttTASNET}. The audio blocks $\vekt{x}_{\ell}$ are encoded into feature vectors $\mat{w}_\ell$. The mask estimation network
produces a sequence of masks from the encoded signal. The masks $\mat{m}_\ell$ are then multiplied with the encoded features $\mat{w}_\ell$ to produce a sequence of output features that are decoded back into the time domain signal by the decoder.
\subsection{Encoder}
The input signal blocks $\vekt{x}_\ell\in\Real^{1\times L_{\mathrm{BL}}}$ are encoded by a 1D convolutional layer with a \ac{ReLU} activation function, $\mathcal{H}_\mathrm{enc}$, such that
\begin{equation}\label{eq:encoder}
 \mat{w}_\ell=\mathcal{H}_\mathrm{enc}\left(\vekt{x}_\ell\mat{B}\right){,}
\end{equation}
where $\mat{B}\in\mathbb{R}^{L_{\mathrm{BL}}\times N}$ is a matrix of trainable weights and $\mat{w}_\ell$ is the encoded feature vector for the $\ell$th signal block. 
\subsection{\texorpdfstring{\Ac{TCN} Mask Estimation Network}{TCN Mask Estimation}}
The mask estimation network 
produces a {mask} $\vekt{m}_\ell$ for {every block $\ell$.} The encoded features {$\mat{w}_\ell$} are first normalized using channelwise layer normalization \cite{ba2016layer}. The normalized features are transformed by a pointwise convolutional layer \cite{convtasnet} which reduces the feature dimension from $N$ to $B$. 
The sequence of features is then processed by a stack of $X$ \acp{CB} with increasing the dilation $f$ of a factor of two per \ac{CB}, i.e. $f\in\{1,\ldots,2^{X-1}\}$. Each \ac{CB} is comprised of a pointwise convolutional 
{layer}, a \ac{PReLU} activation function, \ac{gLN} \cite{convtasnet}, and a depthwise separable convolutional layer \cite{convtasnet}.
The pointwise convolutional layer has $B$ input channels and $H$ output channels. The depthwise separable convolutional layer has $H$ input channels and $B$ output channels. 
The $X$ \acp{CB} of increasing dilation is repeated $R$ times. This {repetition} widens the \ac{RF} of the network to a {lower} degree than continuing to increase the dilation whilst also producing a deeper layered network with more parameters per second of the \ac{RF}. The \ac{RF} of the \ac{TCN}, measured in seconds, is defined as
\begin{equation}
    \label{eq:temporalcontext}
    \hspace*{-0.8ex}\mathcal{R}(L_{\mathrm{BL}},R,X,P)\!=\!
    \frac{L_{\mathrm{BL}}}{2\mathrm{f}_s}
    \!\left(\!1\!+\!{R}(P\!-\!1)\sum_{i=1}^{X}2^{X-i}\!\right)\!
\end{equation}
where $P$ is the kernel size in the \ac{CB} and $\mathrm{f}_s$ defines the sampling rate in Hz.
Proceeding the \acp{CB} is a \ac{PReLU} activation function, followed by a pointwise convolutional layer which transforms the feature dimension from $B$ to $N$. A ReLU activation function is used to produce a set of non negative masks
defined as $\mat{m}_{\ell}\in \Real^{1\times N}$.
\subsection{Decoder}
The decoder is a transposed 1D convolutional layer that decodes the masked encoded mixture ${\mat{v}_{\ell}=\mat{m}_{\ell}\odot\mat{w}_\ell}$ back into the time domain. {The operator $\odot$ denotes the Hadamard product.} The transposed convolutional operation is defined as
\begin{equation}
 \hat{\vekt{s}}_{\ell}=(\mat{m}_{\ell}\odot\mat{w}_\ell)\mat{U}=\mat{v}_{\ell}\mat{U}
\end{equation}
where $\mat{U}\in \Real^{N\times L_{\mathrm{BL}}}$ and $\hat{\mat{s}}_{\ell}$ is the decoded time domain block.
\subsection{Objective Function}
The \ac{SISDR} objective function
\cite{LeRoux} is used for training the \ac{DAE} network. To use \ac{SISDR} as an objective function, the negative \ac{SISDR} is computed such that the network is optimized to maximize the \ac{SISDR} of the estimated speech signal. The \ac{SISDR} loss function is defined as
\begin{equation} 
\label{eq:DefSISDR}
\mathcal{L}_\text{SISDR}(\hat{\vekt{s}},\vekt{s}_\mathrm{dir}): 
= - 10\log_{10} \frac{\left\Vert 
\frac{\langle \hat{\vekt{s}},\vekt{s}_\mathrm{dir}\rangle 
\vekt{s}_\mathrm{dir}}{\Vert \vekt{s}_\mathrm{dir}\Vert^{2}}\right\Vert^{2}}{\left\Vert\hat{\vekt{s}}-\frac{\langle \hat{\vekt{s}},\vekt{s}_\mathrm{dir}\rangle 
\vekt{s}_\mathrm{dir}}{\Vert \vekt{s}_\mathrm{dir}\Vert^{2}}\right\Vert^{2}}.
\end{equation}
%
\section{Data and Experiments}\label{sec:4}
\subsection{Data}
WHAMR \cite{WHAMR}, a monaural noisy reverberant two speaker speech corpus, and an extension of WHAMR, 
{denoted as WHAMR\_ext in the following},
are used for all experiments. Only the first speaker{s'} audio clips are used since the focus is on single speaker dereverberation. The \acp{RIR} are generated using the pyroomacoustics \cite{pyroomacoustics} software framework. RT60 values for the \acp{RIR} are randomly generated between 0.1s and 1s {in WHAMR}. To create WHAMR\_ext, reverberant speech with larger RT60 values between 1s and 3s were simulated following the same routine as {for} WHAMR. 
{Scripts} to recreate WHAMR\_ext {can} be found on github\footnote{{Mixing script available online at} \url{https://github.com/jwr1995/WHAMR\_ext}}.

An 8kHz sample rate is used for all audio. For each corpus, WHAMR and WHAMR\_ext, the training set is comprised of 20,000 speech examples. For training, audio clips are truncated
or padded to 4~seconds, resulting in a total of 22.22 hours of data being used. This approach is used to address signal length mismatches in batches during training \cite{WHAMR}. For validation 5000 audio examples are used {resulting in} 14.65 hours of speech and for testing 3000 audio examples are used{, i.e.~}9 hours of speech. All models are evaluated on the test set.
\subsection{Training Configuration}
All experiments are done using the speechbrain speech processing software framework \cite{speechbrain}. Training is performed over 100 epochs with an initial learning rate set to $10^{-3}$ that is halved if there is no improvement in the average \ac{SISDR} over the validation set after 3 epochs.

The number of blocks, $X$, in the dilated stack of the \ac{TCN} was varied from 1 to 10 and the number of repeats, $R$, of the stack itself was varied from 1 to 8. The  rest of the network's configuration is fixed. The encoder has $L_{\mathrm{BL}}=16$ input channels and $N=512$ output channels. The \ac{TCN} is configured such that there are $B=128$ output channels from the bottleneck layer and each \ac{CB} has $H=512$ internal convolutional channels and a kernel size $P=3$.

\subsection{Evaluation Metrics}
A number of metrics were considered for evaluating the dereverberation performance objectively {\cite{Goetze2}}. \ac{SISDR} is reported for all experiments. In addition \ac{PESQ} \cite{PESQ},
\ac{ESTOI} \cite{estoi} and \ac{SRMR} \cite{srmr} are reported for some models. \ac{PESQ} is an objective measure of speech quality. \ac{ESTOI} is an objective measures of speech intelligibility. \ac{SRMR} is a non-intrusive measure of reverberation energy. $\Delta$-measures show the improvement over the reverberant speech, $x[i]$.
%
\section{Results}\label{sec:5}

\paragraph{ $\Delta$ \ac{SISDR} on WHAMR and WHAMR\_ext}
The $\Delta$~SISDR results for the models trained and evaluated on the WHAMR dataset can be seen in Table \ref{tab:d_sisdr_results:whamr} {for varying $X$ and $R$}. These parameters are varied such that they change the receptive field and model size of \acp{TCN} where $X$ has more effect on the \ac{RF} (cf. Eq. (\ref{eq:temporalcontext})) and both increase the number of layers in the network but $R$ has more effect on the temporal parameter density, i.e. number of parameters per second of the receptive field. Note that one \ac{CB} has ${B H+H+H P+H B=133,120}$~parameters. 

\begin{table}[!ht]
\setlength\tabcolsep{2.5pt}
\caption{ $\Delta$ \ac{SISDR} {in dB}
for all TCN configurations trained and evaluated on WHAMR, best performing model for number of \acp{CB} ($X\cdot R$) in \ac{TCN} shown in bold.} 
\centering
\begin{tabular}{cc|cccccccccc|}
\cline{3-12}
&  & \multicolumn{10}{c|}{\cellcolor[HTML]{C0C0C0} $X$}\\ 
\cline{3-12} 
 &  & 
 \cellcolor[HTML]{C0C0C0}1 & \cellcolor[HTML]{C0C0C0}2 & \cellcolor[HTML]{C0C0C0}3 & \cellcolor[HTML]{C0C0C0}4 & \cellcolor[HTML]{C0C0C0}5 & \cellcolor[HTML]{C0C0C0}6 & \cellcolor[HTML]{C0C0C0}7 & \cellcolor[HTML]{C0C0C0}8 & \cellcolor[HTML]{C0C0C0}9 & \cellcolor[HTML]{C0C0C0}10 \\ \hline
\multicolumn{1}{|c}{\cellcolor[HTML]{C0C0C0}} & \multicolumn{1}{|c|}{\cellcolor[HTML]{C0C0C0}1} & \bfseries 1.88 & \bfseries 2.61 &  \bfseries 3.32 & \bfseries 4.05 & \bfseries 4.66 & \bfseries 5.09 & \bfseries 5.41 & \bfseries 5.65 & \bfseries 5.67 & 5.68 \\
\multicolumn{1}{|c}{\cellcolor[HTML]{C0C0C0}}& \multicolumn{1}{|c|}{\cellcolor[HTML]{C0C0C0}2} & 2.48 & 3.58 & 4.45 & 5.25 & \bfseries 5.92 & \bfseries 6.26 & \bfseries 6.47 & 6.45 & 6.60 & 6.63 \\
\multicolumn{1}{|c}{\cellcolor[HTML]{C0C0C0}}& \multicolumn{1}{|c|}{\cellcolor[HTML]{C0C0C0}3} & 2.95 & 4.08 & 4.94 & 5.94 & 6.43 & \bfseries 6.80 & \bfseries 6.88 & 6.94 & 7.02 & 7.01 \\
\multicolumn{1}{|c}{\cellcolor[HTML]{C0C0C0}} & \multicolumn{1}{|c|}{\cellcolor[HTML]{C0C0C0}4} & 3.28 & 4.46 & 5.47 & \bfseries 6.53 & \bfseries 6.97 & \bfseries 7.01 & \bfseries 7.16 & \bfseries 7.23 & 7.14 & 7.11 \\
\multicolumn{1}{|c}{\cellcolor[HTML]{C0C0C0}} & \multicolumn{1}{|c|}{\cellcolor[HTML]{C0C0C0}5} & 3.54 & 4.82 & 5.86 & 6.70 & \bfseries 7.06 & \bfseries 7.31 & 7.29 & 7.32 & 7.42 & 7.44 \\
\multicolumn{1}{|c}{\cellcolor[HTML]{C0C0C0}} & \multicolumn{1}{|c|}{\cellcolor[HTML]{C0C0C0}6} & 3.74 & 4.99 & 6.16 & 6.87 & 7.25 & \bfseries 7.37 & 7.45 &  7.51 & 7.47 & 7.4{0} \\ 
\multicolumn{1}{|c}{\multirow{-8}{*}{\cellcolor[HTML]{C0C0C0}$R$}} & \multicolumn{1}{|l|}{\cellcolor[HTML]{C0C0C0}8} & 4.09 & 5.55 & 6.44 & 7.12 & \bfseries 7.44 & \bfseries 7.63 & 7.59 & 7.54 & 7.48 & 7.4{0} \\ \hline
\end{tabular}
\label{tab:d_sisdr_results:whamr}
\end{table}

{Results in Table~\ref{tab:d_sisdr_results:whamr}} 
show that for smaller models ($\lesssim2$M parameters)  it is preferable to have a larger \ac{RF}, i.e.{~a higher} $X$ value, than a deeper network per the \ac{RF}, i.e.{~a higher} $R$ value. For example, for a model with 12 \acp{CB} the best performing model configuration is $(X,R)=(6,2)$ where $X=6$ is the largest possible value for the 4 possible \ac{TCN} configurations, $\{(X,R)\}=\{(4,3),(3,4),(6,2),(2,6)\}$. 
The importance of widening the receptive field is also apparent in the first row of Table \ref{tab:d_sisdr_results:whamr} where $R=1$ remains constant for best performance for the first 9 \acp{CB}. This importance of having $X>R$ disappears as the number of \acp{CB} surpasses 36 ($X=6, R=6$) at which point the best performance is gained by models with $R>X$, i.e.~more importance is given to a deeper network than a wider receptive field.

\begin{table}[!ht]
\setlength\tabcolsep{2.5pt}
\caption{ $\Delta$ \ac{SISDR} 
{in dB}
for all TCN configurations trained and evaluated on WHAMR\_ext, best performing model for number of \acp{CB} ($X\cdot R$) in \ac{TCN} shown in bold.}
\centering
\begin{tabular}{cc|cccccccccc|} 
\cline{3-12}
&  & \multicolumn{10}{c|}{\cellcolor[HTML]{C0C0C0}$X$}\\ 
\cline{3-12} 
 &  & 
 \cellcolor[HTML]{C0C0C0}1 & \cellcolor[HTML]{C0C0C0}2 & \cellcolor[HTML]{C0C0C0}3 & \cellcolor[HTML]{C0C0C0}4 & \cellcolor[HTML]{C0C0C0}5 & \cellcolor[HTML]{C0C0C0}6 & \cellcolor[HTML]{C0C0C0}7 & \cellcolor[HTML]{C0C0C0}8 & \cellcolor[HTML]{C0C0C0}9 & \cellcolor[HTML]{C0C0C0}10 \\ 
 \hline
\multicolumn{1}{|c}{\cellcolor[HTML]{C0C0C0}}& \multicolumn{1}{|c|}{\cellcolor[HTML]{C0C0C0}1} & \bfseries 3.04 & \bfseries 3.85 & \bfseries 4.67 & \bfseries 5.79  & \bfseries 6.84  & \bfseries 7.68  & \bfseries 8.09  & \bfseries 8.55  & \bfseries 8.69  & 8.69  \\
\multicolumn{1}{|c}{\cellcolor[HTML]{C0C0C0}}& \multicolumn{1}{|c|}{\cellcolor[HTML]{C0C0C0}2} & 3.65 & 4.76 & 6.11 & 7.44  & 8.56  & \bfseries 9.19  & \bfseries 9.52  & \bfseries 9.64  & 9.76  & 9.79  \\
\multicolumn{1}{|c}{\cellcolor[HTML]{C0C0C0}}& \multicolumn{1}{|c|}{\cellcolor[HTML]{C0C0C0}3} & 4.06 & 5.44 & 6.98 & 8.42  & 9.29  & \bfseries 9.83  & \bfseries 10.13 & \bfseries 10.19 & \bfseries 10.21 & 10.15 \\
\multicolumn{1}{|c}{\cellcolor[HTML]{C0C0C0}}& \multicolumn{1}{|c|}{\cellcolor[HTML]{C0C0C0}4} & 4.45 & 6.10  & 7.62 & 8.96  & 9.68  & 10.11 & \bfseries 10.41 & \bfseries 10.42 & 10.42 & 10.47 \\
\multicolumn{1}{|c}{\cellcolor[HTML]{C0C0C0}}& \multicolumn{1}{|c|}{\cellcolor[HTML]{C0C0C0}5} & 4.70  & 6.51 & 8.21 & 9.36  & 10.01 & 10.37 & \bfseries 10.60  & \bfseries 10.62 & 10.54 & 10.50  \\
\multicolumn{1}{|c}{\cellcolor[HTML]{C0C0C0}}& \multicolumn{1}{|c|}{\cellcolor[HTML]{C0C0C0}6} & 4.96 & 6.85 & 8.48 & 9.63  & 10.15 & 10.49 & \bfseries 10.74 & 10.77 & 10.67 & 10.60  \\
\multicolumn{1}{|c}{\cellcolor[HTML]{C0C0C0}}& \multicolumn{1}{|c|}{\cellcolor[HTML]{C0C0C0}7} & 5.29 & 7.14 & 8.75 & 9.71  & 10.34 & 10.61 & 10.72 & 10.68 & 10.76 & 10.70  \\
\multicolumn{1}{|c}{\multirow{-8}{*}{\cellcolor[HTML]{C0C0C0}$R$}} & \multicolumn{1}{|c|}{\cellcolor[HTML]{C0C0C0}8} & 5.45 & 7.44 & 9.03 & 10.02 & 10.49 & \bfseries 10.80  & \bfseries 10.81 & 10.67 & 10.68 & 10.57 \\ \hline
\end{tabular}
\label{tab:d_sisdr_results:whamr_ext}
\end{table}

Comparing the best performing models for increasing the number of \acp{CB} ($X\cdot R$) trained on WHAMR ({cf.~}Table~\ref{tab:d_sisdr_results:whamr}) and WHAMR\_ext ({cf.~}Table~\ref{tab:d_sisdr_results:whamr_ext}) indicates that for {a dataset containing} only larger RT60 values between 1s and 3s it is more preferable to increase the model's \ac{RF} as the number of \acp{CB} in the \ac{TCN} increases up to 42{, i.e.~$(X,R)=(7,6)$}. 

For RT60 values between 0.1s and 1s {(WHAMR corpus, Table~\ref{tab:d_sisdr_results:whamr})} the model only benefits from putting more emphasis when expanding the \ac{RF} up to 36 \acp{CB}. 

\begin{table*}[h]
\setlength\tabcolsep{4pt}
\centering
\begin{tabular}{|c|c|cc|cc|cc|cc|cc|cc|}
\hline
\bfseries \cellcolor[HTML]{C0C0C0}train & \bfseries \cellcolor[HTML]{C0C0C0}eval &  \cellcolor[HTML]{C0C0C0}$X$ & \cellcolor[HTML]{C0C0C0}$R$ & \bfseries \cellcolor[HTML]{C0C0C0}\# params & \cellcolor[HTML]{C0C0C0}$\mathcal{R}$ \bfseries (s) & \cellcolor[HTML]{C0C0C0}\bfseries \ac{SISDR} & \cellcolor[HTML]{C0C0C0} \bfseries $\Delta$ \ac{SISDR} & \cellcolor[HTML]{C0C0C0} \bfseries PESQ & \cellcolor[HTML]{C0C0C0} \bfseries $\Delta$ PESQ & \cellcolor[HTML]{C0C0C0} \bfseries ESTOI & \cellcolor[HTML]{C0C0C0} \bfseries $\Delta$ ESTOI & \cellcolor[HTML]{C0C0C0} \bfseries \ac{SRMR} & \cellcolor[HTML]{C0C0C0} \bfseries $\Delta$~SRMR \\ \hline
WHAMR & WHAMR & 6 & 8 & 6.6M & 1.02 & 12.03 & 7.63 & 3.46 & 0.91 & 0.93 & 0.15 & 8.7 & 2.26 \\ \hline
WHAMR & WHAMR\_ext & 8 & 8 & 8.8M & 4.09 & 5.89 & 9.64 & 2.3 & 0.94 & 0.74 & 0.35 & 8.48 & 5.72 \\ \hline
WHAMR\_ext & WHAMR & 6 & 8 & 6.6M & 1.02 & 10.79 & 6.39 & 3.24 & 0.69 & 0.92 & 0.14 & 8.81 & 2.36 \\ \hline
WHAMR\_ext & WHAMR\_ext & 7 & 8 & 7.7M & 2.04 & 7.07 & 10.81 & 2.46 & 1.11 & 0.81 & 0.42 & 9.18 & 6.42 \\ \hline
\end{tabular}
\caption{Best performing models for models trained on WHAMR and WHAMR\_ext evaluated on each test set.}
\label{tab:best_performing}
\end{table*}
Table~\ref{tab:best_performing} shows the best performing models in terms of SISDR, trained on each training set and evaluated on each test set. These results show that performance improvements in \ac{SISDR} are similarly replicated in \ac{SRMR} \ac{PESQ} and \ac{ESTOI} measures and thus correspond to an objective improvement in {perceived reverberation,} quality and intelligibility of speech. 
Tables~\ref{tab:d_sisdr_results:whamr} and \ref{tab:d_sisdr_results:whamr_ext} have been calculated also for \ac{SRMR}, \ac{PESQ} and \ac{ESTOI}, which show similar trends. Table~\ref{tab:best_performing} also shows that when evaluated on WHAMR the model that was trained on WHAMR\_ext gives better dereverberation performance (in \ac{SRMR}) but was more distorted (in \ac{SISDR}), c.f. \ac{SISDR} and \ac{SRMR} results in rows 1 \& 3. It is however expected that training on both WHAMR and WHAMR\_ext jointly would lead to further generalisation improvements on the WHAMR test set.

\begin{figure}[h]
    \centering
    \begin{subfigure}[b]{\columnwidth}
        \centering
        \includegraphics[width=\columnwidth]{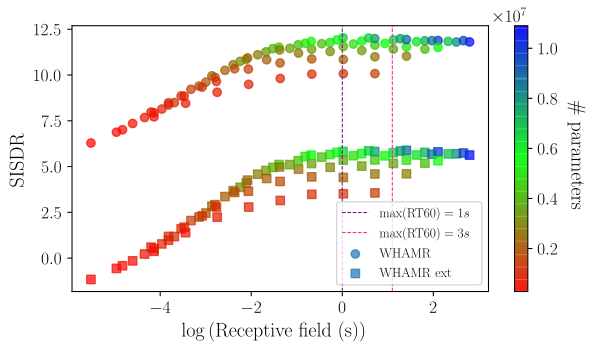}\\[-1.5ex]
        \caption{ Trained on WHAMR.}
        \label{fig:sisnr:a}
    \end{subfigure}\\[-0.5ex]
    \begin{subfigure}[b]{\columnwidth}
        \centering
        \includegraphics[width=\columnwidth]{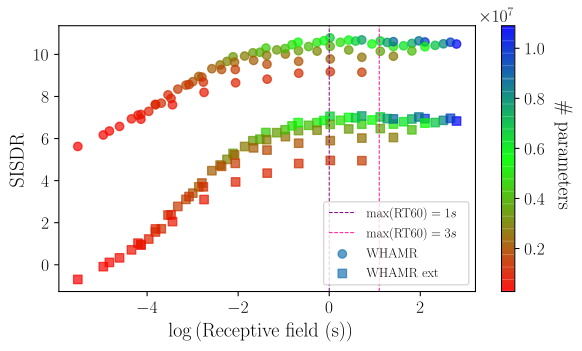}\\[-1.5ex]
        \caption{ Trained on WHAMR\_ext.}
        \label{fig:sisnr:b}
    \end{subfigure}\\[-0.5ex]
    \caption{SISDR depending on logarithmic \ac{RF}. Circles and squares indicate evaluation on WHAMR and WHAMR\_ext test sets, respectively. Maximum RT60s of WHAMR and WHAMR\_ext are shown by dashed lines. Colour scale indicates the number of model parameters.}
    \label{fig:sisnr}
\end{figure}
\begin{figure}
    \centering
    \includegraphics[width=\columnwidth]{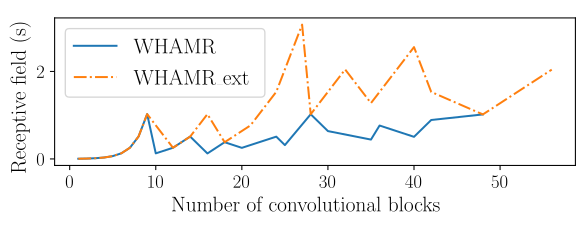}
    \caption{\acp{RF} for the best performing models in Tables \ref{tab:d_sisdr_results:whamr} and \ref{tab:d_sisdr_results:whamr_ext} shown by increasing model size measured in number of \acp{CB} (one \ac{CB} is 133,120 parameters). Line colour and style indicates the training and test set used.}
    \label{fig:rfs_cbs}
\end{figure}

\paragraph{\ac{RF} and model size}
Figure \ref{fig:sisnr} shows the results for models trained on WHAMR (upper panel) and WHAMR\_ext (lower panel) depending on the \ac{RF} and model size. Note that the \ac{SISDR} measure is used here, as opposed to the $\Delta$ \ac{SISDR} measure, because it is later compared with \ac{SRMR} in this section. 
In terms of model size, the \ac{SISDR} performance that can be achieved by TCNs alone saturates as the number of parameters approaches 
6M, for both WHAMR and WHAMR\_ext when evaluated on WHAMR. For evaluation on WHAMR\_ext, the \ac{SISDR} performance also saturates as it approaches 6M parameters but the results for the model trained on WHAMR in Table \ref{tab:best_performing} indicate it may benefit from the larger model size when having to dereverb larger reverberation times especially when trained only with smaller RT60s.
Figure \ref{fig:sisnr} further illustrates that the \ac{SISDR} performance saturates before the \ac{RF} reaches 1s for models trained on WHAMR, i.e. the highest occurring RT60 in WHAMR. The \Acp{RF} of the best performing models can be seen in Table \ref{tab:best_performing}. The best model trained and evaluated on WHAMR in terms of \ac{SISDR} has an \ac{RF} of 1.02s. Analysing the same models but evaluated on WHAMR\_ext the best \ac{SISDR} performance is attained with an \ac{RF} of 4.09s. 
For models trained on WHAMR\_ext the optimal model evaluated on WHAMR has an \ac{RF} of 1.02s and an \ac{RF} of 2.04s when evaluated on WHAMR\_ext.
Figure \ref{fig:rfs_cbs} analyses the best performing models on each of the two datasets by their \Acp{RF} depending on model size (i.e. number of \Acp{CB}).
This indicates that for larger RT60 values \acp{TCN} benefit from having a larger \ac{RF} as most of the best performing models evaluated on WHAMR\_ext have a larger \ac{RF} than the best performing models evaluated on WHAMR.
The \ac{SISDR} of the estimated signal improves even with RT60s larger than the \ac{RF} as can be seen in Figure~\ref{fig:sisnr}. This implies the network learns how to estimate masks that suppress the characteristic of reverberation, as opposed to trying to perform a blind convolution operation based on an estimated \ac{RIR} representation in the network. As such, it should be possible to dereverb \Acp{RIR} of any RT60 value regardless of \ac{RF} but Table~\ref{tab:best_performing} indicates that having an \ac{RF} close to the maximum RT60 value is optimal.
%

\paragraph{\ac{SRMR} vs \ac{SISDR}}
\begin{figure}[h]
    \centering
    \begin{subfigure}{\columnwidth}
        \centering
        \includegraphics[width=\columnwidth]{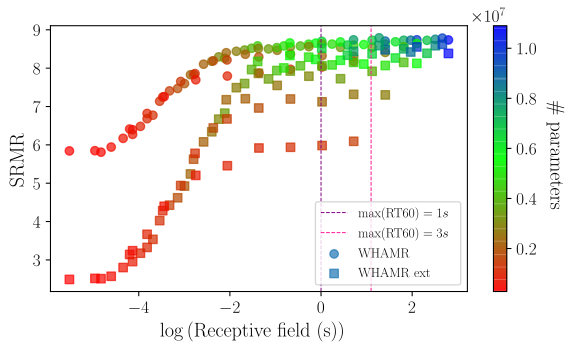}\\[-1.5ex]
        \caption{ Trained on WHAMR.}
         \label{fig:srmr:a}
    \end{subfigure}\\[-0.5ex]
    \begin{subfigure}{\columnwidth}
        \centering
        \includegraphics[width=\columnwidth]{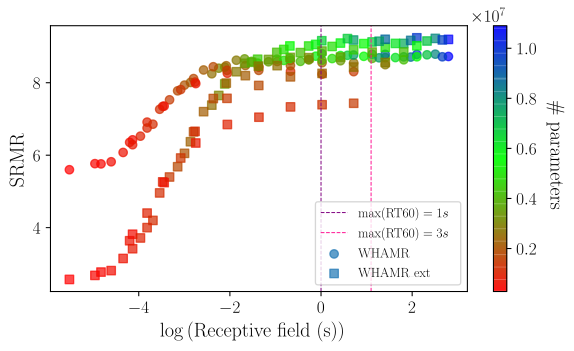}\\[-1.5ex]
        \caption{ Trained on WHAMR\_ext.}
         \label{fig:srmr:b}
    \end{subfigure}\\[-0.5ex]
    \caption{SRMR depending on logarithmic \ac{RF}. Circles and squares indicate evaluation on WHAMR and WHAMR\_ext test sets, respectively. Maximum RT60s of WHAMR and WHAMR\_ext are shown by dashed lines. Colour scale indicates the number of model parameters.}
    \label{fig:srmr}
\end{figure}
Comparing results of \ac{SISDR} (Figure~\ref{fig:sisnr}) with \ac{SRMR} (Figure \ref{fig:srmr}) indicates that with sufficiently large model sizes ($\gtrsim4$M parameters) much of the residual distortions in the signal are artifacts introduced by the network and not reverberation itself. This also indicates that the larger the reverberation time, the more residual distortions are present in the estimate clean speech signal. Audibly, the distortions present in RT60s $\lesssim1$ were not particularly noticeable however as the RT60 approaches 3s they become more noticeable as informal listening tests showed. The performance of \ac{SRMR} also saturates slightly earlier than that of \ac{SISDR} similarly implying that some of the gain from increasing the model size has more correlation to reducing artefact distortions in $\hat{s}[i]$ than any residual reverberant effects.
This is an argument in favour of using the \ac{SISDR} function over a pure reverberation based measure like \ac{SRMR} as the loss function because \ac{SRMR} is more agnostic to general distortions in the signal.
%
\paragraph{Improvement on WHAMR vs WHAMR\_ext}
\begin{figure}[h]
    \centering
    \includegraphics[width=\columnwidth]{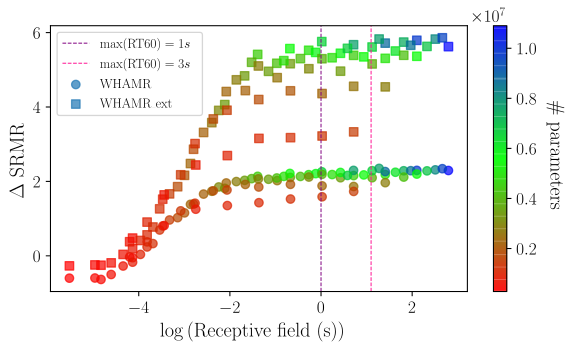}
    \caption{$\Delta$~SRMR for model trained on WHAMR depending on logarithmic \ac{RF}. Circles and squares indicate evaluation on WHAMR and WHAMR\_ext test sets, respectively. Maximum RT60s of WHAMR and WHAMR\_ext are shown by dashed lines. Colour scale indicates the number of model parameters.}
    \label{fig:srmr_d}
\end{figure}
The $\Delta$~\ac{SRMR} results in Figure \ref{fig:srmr_d} demonstrate that greater reverberation improvement can be attained on the more reverberant WHAMR\_ext dataset but that larger model sizes ($\gtrsim4$M parameters) are required to fully capitalize on this. Also for both WHAMR\_ext evaluation there is much broader distribution of values as the receptive field increases. This is related to the $R$ variable in the TCN, in other words using a deeper network leads to a more significant improvement when evaluating \ac{SRMR} improvement on larger RT60s.
%
\section{Conclusion}\label{sec:6}
In this paper \ac{TCN}s were analysed for their application in dereverberation tasks.
It was found that for smaller models more emphasis should be put on widening the \ac{RF} of the network than using more layers in a network.
The model performance in both \ac{SISDR} and \ac{SRMR} starts to saturate around a model size of $4$M parameters. 
It was shown that the for larger RT60 values there is strong evidence that having a wider receptive field is important for achieving optimal performance. This is especially true when the model is trained on smaller RT60s. 
It was found that \ac{SRMR} performance was fairly agnostic to the variation in RT60 values but \ac{SISDR} performance was significantly impacted.
This indicates that much of the distortion remaining in the signal maybe modelling errors as opposed to reverberant effects.

\bibliographystyle{IEEEtran} 
\bibliography{bibliography}

\end{document}